\documentstyle[eqsecnum,aps]{revtex}

\begin{document}

\preprint{}
\title{Electronic structure of strained InP/GaInP quantum dots.
}
\author{Craig Pryor\cite{email},  M-E. Pistol, L. Samuelson,}
\address{
Department of Solid State Physics \\
Box 118, Lund University\\
S-221 00 Lund, Sweden
}
\maketitle
\begin{abstract}
We calculate the  electronic structure of nm scale 
InP islands embedded in ${\rm Ga}_{0.51}{\rm In}_{0.49}{\rm P}$.
The calculations are done
in the envelope approximation and include the effects of strain, piezoelectric
polarization, and mixing among 6  valence bands. The
electrons are confined within the entire island, while the holes are
 confined to
strain induced pockets. One pocket forms a ring at the bottom of the
island near the substrate interface, while the other is above the island
 in the 
GaInP.  The two sets of hole states are decoupled. 
Polarization dependent dipole matrix elements are calculated for both
types of hole states. 
\end{abstract}
\pacs{
}


\section{Introduction}
In recent years great progress has been made 
in the fabrication and measurement
of  semiconductor  quantum dots made by Stranski-Krastonov growth.
In this technique, material is deposited epitaxially onto a substrate 
that is lattice
 mismatched
to the deposited material.  Due to the mismatch, the deposited material 
spontaneously
forms nm scale islands, the size and shape of  which are both material and 
growth-condition dependent. 
The pyramidal shape
and presence of strain in the islands makes for a potentially rich 
electronic structure.
Theoretical studies of strained islands have employed
various degrees of approximation to the geometry, strain distribution, and
electron dynamics,  ranging from single band models of 
hydrostatically strained islands, to multiband models including 
realistic shapes and strain distributions \cite{bastard,grundman,cusack,fu}.

In this paper we consider  InP islands buried in
${\rm Ga}_{0.51}{\rm In}_{0.49}{\rm P}$ (GaInP)\cite{TEM}, 
a material combination that has so far
received little theoretical attention.
The geometry is made as realistic as possible by adopting the shape
 shown in
 Fig. 1, which has been
observed using atomic force microscopy and transmission
electron microscopy \cite{TEM}. 
We assume that different size islands have the same shape as shown
in Fig. 1, but are simply rescaled. We parameterize the size of the island by
 the height
in the $\hat z$ direction, $h$. 
In most of the calculations we set
 $h=15 ~\rm nm$, but we will consider variations as well.
Note that the wetting layer is not included in our calculation.
The wetting layer does not  significantly alter the  strain distribution in
the island
because it is pseudomorphically strained. Also, its measured thickness
is on the order of the grid spacing used, and
hence its inclusion in the electronic structure portion of the calculation
 would
be unjustified.
The calculation  is done in two steps:
the strain distribution is determined using continuum elastic theory, and
then used in a strain dependent $\bf k\cdot p$ Hamiltonian.

\section{Strain}

The strain  is
computed  by minimizing the
free energy for a cubic crystal, \cite{LL}\\
\begin{eqnarray}
F=\int d^3x {1\over 2}&C&_{xxxx}(\sigma_{xx}^2+\sigma_{yy}^2+
\sigma_{zz}^2)+\cr
&C&_{xxyy}(\sigma_{xx} \sigma_{yy} + \sigma_{xx} \sigma_{zz} +
	 \sigma_{yy}\sigma_{zz})+\cr
    &2&C_{xyxy}(\sigma_{xy}^2+\sigma_{xz}^2+\sigma_{yz}^2)
    -\alpha(\sigma_{xx}+\sigma_{yy}+\sigma_{zz}), \label{eq:F}
	\end{eqnarray}
where $\sigma_{ij}$ is the strain tensor, 
 the $C$'s are material depended elastic constants, and $\alpha$ is a
parameter
used to enforce the lattice mismatch between the two materials.
The strain is given in terms of the displacement by
 $\sigma_{ij}={1\over 2}(\partial_i u_j + \partial_j u_i)$.
$F$ is constructed as a function of  the $u_i$'s
on a cubic grid with  derivatives replaced by differences,
and then $F$ is minimized  using the conjugate gradient algorithm.
The grid consisted of a periodic box of $130\times 130 \times 120$ sites
with the island contained within a $65\times 65\times 20$ region. The grid
is reduced  $1/4$ by exploiting the  symmetries of the system.
The size of the box is chosen sufficiently large to make the band energies
within $1\%$ of the bulk GaInP values at the points furthest from the
island. Errors due to the finite system should be $\le 1\%$ at the island.
 No constraints need to be 
directly imposed on the minimization since far from the island the
strain has relaxed to its bulk ($=0$) value.

The lattice mismatch between the two materials is implemented as follows.
The grid is chosen to be commensurate with the barrier material
so the grid corresponds to some multiple of the barrier material's
{\it crystal} unit cell.
Then for unstrained barrier material we would have
$u_{i}=0$.
But  this means that the grid is
wrong for the island material since it has a different crystal unit
cell. 
Unstrained island material should actually have
a nonzero $u_{i}$ to reflect the fact that its crystal unit cell
has a different size. 
If the island and barrier lattice constants are $a_I$ and $a_B$ respectively,
then setting
$\alpha = (C_{xxxx}+2C_{xxyy})(a_I - a_B) / a_B$ in the island
 causes
the minimum of F to be at
$\sigma^0_{ij} = \delta_{ij} (a_I - a_B) / a_B$, which is the 
desired amount of  strain  in the grid.
Because
$\sigma^0_{ij}$ is completely fictitious, and
reflects only the  straining of  the coordinate system,
the physical strain is obtained by subtracting  $\sigma^0_{ij}$ from the
   $\sigma_{ij}$
that minimizes $F$ for the heterostructure.

There is some freedom in the choice of grid  differences used to approximate
$F$.  Using symmetric differences,  $\partial_x f(x)\rightarrow 
[f(x+h)-f(x-h)]/2a$, 
 produces 
oscillatory solutions that must be smoothed.  Oscillatory solutions are
in fact a common problem with finite difference approximations to equations
involving first derivatives.  For the case at hand this problem is avoided by
evaluating $F$ within each unit cube as a function of the $u_i$'s at the eight
corners.
This is essentially a low-order finite element method. Although more
 sophisticated
approximations may yield more more accurate strains, we are constrained by
the fact that the strain is to be used with the cubic grid which is most
 convenient
for the electronic structure part of the calculation.
Only one strain field is needed for all sizes
 since it is a dimensionless quantity,  and the strain for a different
island size is obtained by a trivial rescaling.

Because the materials lack inversion symmetry the
strain produces a polarization
given by 
$P_i  = e_{ijk} \sigma_{jk}$, leading to an additional electrostatic potential.
  For III-IV semiconductors,
 the only nonzero
elements of the piezoelectric tensor are 	
$ e_{xyz}=e_{zxy}=e_{yzx} \equiv e_{14}$.
From the polarization it is a simple matter to compute the 
electrostatic potential $V_p$ by numerically solving Poisson's equation.
Dipoles are induced at  some of the edges, and on the 
 $\{1 \bar 1 1\}$ and
$\{ \bar 1 1 1 \}$ faces.
Because the islands are primarily under compression the  faces have a negative
charge (Fig. 2).
The piezoelectric effect contributes only a small amount to the
potential felt by carriers in the island, and most of the modification of the
potential is outside the island, as seen in Fig. 2. 
In  studies of InAs islands the piezoelectric potential was important
because the islands were assumed to have four-fold rotational symmetry which
was broken by the piezoelectric potential  \cite{grundman}.
The six-sided InP
islands lack such a symmetry to begin with, and the piezoelectric potential
does not make a qualitative change in the spectrum.  

\section{Hamiltonian}

The electronic structure  is calculated in the  
envelope approximation using $\sigma_{ij}$
and $V_p$ computed above. The electron Hamiltonian is \cite{bahder}
\begin{eqnarray}
H_e=E_c-{\hbar^2\over 2m}\nabla^2 + a_c \sigma  -eV_p
\end{eqnarray}
where $E_c$ is the local conduction-band edge in 
the absence of strain, $a_c$ is
the conduction-band deformation potential, 
and $\sigma ={\rm tr}  \sigma_{ij}$.
For the valence band we used a 6-band model
given by \cite{bahder} \\
\begin{eqnarray}
H_h&=&H_0+H_s-eV_p\\
H_0&=&\left (\matrix{
-P+Q                       & -S^*                    &R               
                  & 0
& \sqrt{3\over2}S         & -\sqrt{2}Q              \cr\cr
-S                             & -P-Q                   &0       
                          & R 
&  -\sqrt{2}R                  & {1\over\sqrt{2}}S \cr \cr
R^*                           & 0                         & -P-Q 
                        & S^*                      
& {1\over \sqrt{2}}S^* & \sqrt{2}R^*             \cr \cr
0                               & R^*                     &S 
                               & -P+Q                    
& \sqrt{2}Q                   & \sqrt{3\over 2}S^* \cr \cr
\sqrt{3\over 2}S^* & -\sqrt{2}R^*       & {1\over \sqrt{2}}S  & \sqrt{2} Q
& -P-\Delta                   & 0                                  \cr \cr
-\sqrt{2} Q              &{1\over \sqrt2}S^* & \sqrt{2}R               
  & \sqrt{3\over 2}S
 & 0                                & -P-\Delta                   \cr \cr
}\right )\\
P&=&-E_v- \gamma_1{\hbar^2\over 2m_0}(\partial_x^2+\partial_y^2+
\partial_z^2)\\
Q&=&-\gamma_2 {\hbar^2 \over 2m_0}(\partial_x^2+\partial_y^2-
2\partial_z^2)\\
R&=&\sqrt{ 3} {\hbar^2 \over 2m_0}\left[ \gamma_2(\partial_x^2-
\partial_y^2)-2i\gamma_3 \partial_x \partial_y \right]\\
S&=&-\sqrt{3} \gamma_3 {\hbar^2\over m_0} \partial_z(\partial_x-
i\partial_y)
\end{eqnarray}
where $E_v$ is the unstrained local valence-band edge, and
 $\gamma_i$ are the Luttinger parameters.
Since the Luttinger parameters vary spatially, we used the prescription
$\gamma_i \partial_x^2 \rightarrow \partial_x \gamma_i \partial_x$.

The strain dependent coupling is given by \cite{bahder}
\begin{eqnarray}
H_s&=&\left (\matrix{
-p+q     & -s^*      &r         & 0                        
 & \sqrt{3\over2}s         & -\sqrt{2}q              \cr\cr
-s                             & -p-q                   &0        & r 
&  -\sqrt{2}r                  & {1\over\sqrt{2}}s \cr \cr
r^*                           & 0       & -p-q      & s^* 
& {1\over \sqrt{2}}s^* & \sqrt{2}r^*             \cr \cr
0                               & r^*          &s      & -p+q 
& \sqrt{2}q                   & \sqrt{3\over 2}s^* \cr \cr
\sqrt{3\over 2}s^* & -\sqrt{2}r^*       & {1\over \sqrt{2}}s  & \sqrt{2} q 
& -a_ve                             & 0                              \cr \cr
-\sqrt{2} q              &{1\over \sqrt2}s^* & \sqrt{2}r  & \sqrt{3\over 2}s 
& 0                                & -a_ve                        \cr \cr
}\right )
\end{eqnarray}
\begin{eqnarray}
p&=&a(\sigma_{xx}+\sigma_{yy}+\sigma_{zz})\\
q&=&b\left[ \sigma_{zz}-{1\over 2} (\sigma_{xx}+\sigma_{yy} ) \right] \\
r&=&{\sqrt{3}\over 2}~b~(\sigma_{xx}-\sigma_{yy}) -id\sigma_{xy}      \\
s&=&-d (\sigma_{xz} - i\sigma_{yz}  )
\end{eqnarray}
where  $a_v$, $b$ and $c$ are
the  hydrostatic and two shear   deformation potentials.
The four band  model may be obtained by taking $\Delta \rightarrow \infty$.
\\

\section{Material Parameters}

The values used for the various material parameters are given in table 1.
These vary considerably in the accuracy with which they have been
measured, with the deformation potentials suffering the most uncertainty.
There is a paucity of data on GaInP, and  most quantities are determined
by interpolating between InP and GaP.  The conduction-band effective
mass in GaInP is simply set to the value for InP, since interpolation is
complicated by the fact that GaP is indirect. 
The hydrostatic deformation potentials 
for  GaInP,
$a_c$ and $a_v$,
are estimated by assuming $a_c/a_v$ is the same as for InP, while
$a_g = a_v + a_c$ is taken from GaInP measurements  \cite{DP,dp2}. 
In computing $V_p$, the dielectric constant is assumed to
take on the InP value throughout the system.

One parameter that bears special mention is
the unstrained valence-band offset. (i.e. the InP valence-band 
energy referenced to that of the GaInP, in the absence of strain.) The
value used is based on transition-metal impurity spectra, and is in agreement
with the value based on Au schottky barrier data \cite{offset}.
The idea is that transition-metal impurities are empirically observed to
have energy levels fixed with respect to the vacuum, relatively independent
of their host environment.
 Thus, by comparing   band edges referenced
to the   impurity levels in two different materials one deduces the relative
 band offsets if the strain could be turned off.
  The  ground state energies of Fe impurities are  
 $0.785 ~\rm eV$  and $0.74 \rm eV$
above the valence band in
InP and GaInP respectively \cite{LB}, so the InP valence band is $45 \rm meV$
below  GaInP.\\

\section{Electronic  Structure}

Before  solving Schr\"{o}dinger's equation
it is instructive to examine the band structure shown in 
Fig. 3.  
The conduction-band is rather ordinary, with electrons confined to the island
by a barrier of about $250 \rm meV$
at the InP/GaInP interface.  Strain reduces the barrier height, but since
most of the band
offset resides in the conduction-band, there is still a substantial barrier.
There is also some modification of the band edge inside the island, but
for the most part it looks like a particle in a box.
The barrier height of $250 \rm meV$ is in good agreement with measurements
using deep-level transient spectroscopy, which give a barrier heigth of 
$240 \rm meV$ \cite{anand}. 

The valence band has a very complex structure. 
Note that  in the absence of strain the island would actually be a weak 
antidot since the 
InP unstrained valence-band offset is $-45 \rm meV$;
The hole confinement is due entirely to strain.
The valence-band edge has peaks near the bottom corners of
the island which extend around the base  in a ring.
The highest points in this ring lie near the \{111\}  planes.
Because of the shallowness of the hole potential, the piezoelectric potential
has a greater impact on the holes than on the electrons.
Looking along the 
 $\hat z$ axis we see the valence band
  also has peaks  in the GaInP immediately above and below the island.

The confined state
energies and wave functions are found by diagonalization of 
a finite difference version of the Hamiltonian
using the Lanczos algorithm.
This is done on the same
cubic grid used to compute the strain,  the only difference being that it
is truncated to
exclude unnecessary regions of barrier material.
The symmetry of the island is not used to reduce the grid.

The first few conduction-band
 wave functions are shown in Fig. 4. These states are
relatively simple to understand if one crudely approximates
the truncated pyramid as a flat box which is smallest in the $\hat z$
direction. The low-lying states all correspond to  excitations in the
$\hat x$ and $\hat y$ directions. One manifestation of the strain is
 the ordering
of the first and second excited states which appears reversed from the naive
expectation. If the island were a simple box, 
the state with a node along the  $[1\bar 1 0]$  direction would be higher 
since the
island is shorter in that direction. However, strain modifies the the
 potential,
 resulting in a different ordering of states.

The valence-band states have a more complex structure, as seen
in Fig. 5.
The states fall into two categories:  states localized near the bottom
of the island, labelled  $A_n$, and those localized above the island in the
 GaInP,
labelled $B_n$.
  The ground state  is $A_0$,  which is peaked around the
 band-edge maxima near the  \{111\} planes.  
  Excited states (lower  valence-band energy) extend around the
ring at the base of the island. 
The first type $B$ state, $B_0$, appears   $15 \rm meV$ from the ground state.
Although the band diagram shows a  pocket below the island,
there are no localized states there.
The localized valence-band states provide 
a partial explanation for the observation
of multiple lines seen in photoluminescence (PL) \cite{apl}. 
An excited hole that relaxed into a $B$ state would be unable to transfer into
one of the $A$ states,  and would instead recombine with
an electron in its ground state.
Examination of the valence-band wave functions shows
that including the split-off band is not a superfluous addition.
The split off component  is typically of order
0.2 of the largest component. 

Since island size depends on growth conditions it is interesting
to examine the dependence of energy on island size, as shown
in Fig. 6.
The conduction-band spacings are
of order $10 \rm meV$ for a height of $15 ~\rm nm$,
increasing to $20 \rm meV$ for a $10~\rm  nm$ island. 
The hole spacings are much smaller due to their higher
effective mass, and the dependence on $h$ is more complex
due to the odd shaped potential. 
It should be noted that although the spacings re not monotonic in $h$, the
individual energies are.

\section{Transitions}
Knowledge of transition rates are important for determining which states will
be experimentally accessible. In order for  the  $B$ states to produce
an observable PL line
 the $B \rightarrow A$ transition rate must be 
sufficiently slow that recombination occurs before relaxation
into an $A$ state. 
A simple method of analysis is to compare the relaxation rates for 
$B_0 \rightarrow A_n$
with  $A_n \rightarrow A_{n-1}$, where $A_n$ is the $A$-state closest in
 energy  
into which the  $B_0$ state could relax.
If, for example, we assume the transition is due to emission of an 
acoustic phonon,
the rate is given by
$T_{i\rightarrow f}  \propto | \langle f|  \exp (i\vec q \cdot \vec r ) |i 
\rangle |^2.$
The  calculated wave functions give
$T_{B_0\rightarrow A_n} {\ \lower-1.2pt\vbox{\hbox{\rlap{$<$}
\lower5pt\vbox{\hbox{$\sim$}}}}\ }  10^{-12} T_{A_n\rightarrow A_{n-1}} $
for $\vec q$ throughout the Brillouin zone. 
This suppression is due entirely to the  highly localized nature of the states,
so a comparable suppression will be obtained for other relaxation processes.

Since multiple PL lines are observed experimentally, it behooves us
to examine optical transitions involving excited states.
The strength and polarization dependence of the radiative transitions
are found by computing the
dipole  matrix elements. These are computed 
by decomposing the envelope wave functions into
linear combinations of the states $\langle X|$,  $\langle Y|$,  $\langle Z|$,
 and
 $\langle S|$, and using the fact that only 
$\langle X | p_x|S\rangle = 
  \langle Y | p_y|S\rangle = 
  \langle Y | p_z|S\rangle$
 is nonzero.
The matrix elements are calculated for light  in the $\hat z$ direction
for both possible polarizations. The results are shown
in Fig. 7, where $I_x$ and $I_y$ are the squares of the
dipole matrix elements for light polarized in the $\hat x$ and $\hat y$ 
directions respectively.
 Casual examination of the wave functions might lead one to believe
that the transition involving the $B$ state would be suppressed since it is
type II. This concern is seen to be unfounded, and
the transition rate from the  $B$ state is comparable to that from the
 $A$ state.
The $B$ transitions  are highly suppressed for  conduction-band
states that are antisymmetric about the [110] direction. The $A$ transitions
are less selective, but still show strong polarization dependence.
The most notable feature of transitions between the conduction-band
ground state and the valence-band states $A_n$ is
that they
are predominantly polarized in the $\hat x$ direction. Therefore, even 
if there is
mixing of the $A$ states, the resulting transition should still be polarized 
in the
$\hat x$ direction.

\section{Conclusions}

Comparison with PL measurements shows good agreement  with
the major features\cite{apl}. 
PL of single quantum dots shows energies in the
range $1.62 -1.64 \rm eV$, which corresponds to  the calculated energies
for island heights in the range  $14 - 17 ~\rm nm$.
The calculated PL spacing of $15 \rm meV$ for $h=15 ~\rm nm$ is in agreement
with the observed multiple PL lines which are $10 - 20 \rm meV$ apart.
Experiments  on single islands
show as many as four lines, which the current
model is unable to explain.  
It should be noted, however, that macroscopic PL measurements which
necessarily average over a range of island sizes and shapes, typically
show two peaks spaced  $15-20 \rm meV$ apart.
Therefore   the dominant characteristic of the spectrum is
consistent with the valence-band double-well potential. The additional
lines appear to be due to some further structure.
The extra structure could be due to asymmetry of the islands, which
would each split  each of the $A$ and $B$ states 
into two localized states, giving a total of four lines.
While appealing, this model is difficult to reconcile with the
observation that the islands have a high degree of symmetry \cite{TEM}.
 
In summary, we have shown that InP islands have a rich electronic structure,
with holes confined to multiple pockets in and around the island.
  The energy levels are consistent with measurements on
single dots, and the calculated structure can produce two PL lines
$15 \rm meV$ apart. While the predicted doubling of PL lines provides an
important existence proof, more work is needed to fully explain the puzzling
proliferation of lines seen in experiments.

\break
\begin{table}
\caption{Material parameters. Unless otherwise noted,  InP values are taken
from Reference 11, and GaInP values are interpolated between InP and GaP values from
Reference 11.
\label{table1}}
\begin{tabular}{lcr}
Parameter&InP&GaInP\\
\tableline
$m_c$    & $0.077 ~m_0$    & $0.077~ m_0$ ~\tablenotemark[1] \\
$\gamma_1$            & $4.95$                                 &  $5.24$  \\
$\gamma_2$            & $1.65$                                 & $1.53$   \\
$\gamma_3$            & $2.35$                                 & $2.21$  \\
$E_g$      & $1.424 ~\rm eV$  &    $1.97 ~\rm eV$~\tablenotemark[2]\\
$\Delta$     & $0.11 ~\rm eV$    & $0.095~ \rm eV$        \\
$a_{gap}$    & $-6.6 ~\rm eV$  & $-7.1 ~\rm eV$~\tablenotemark[2] \\
$a_c$                       & $-7.0 ~\rm eV$~\tablenotemark[3] & $-7.5 ~\rm 
eV~\tablenotemark[4] $            \\
$a_v$     & $0.4~ \rm eV$    & $0.4~ \rm eV$~\tablenotemark[4] \\
 $b$         & $-2.0 ~\rm eV$    &  $-1.9~ \rm eV$ \\
$d$        &  $-5.0~ \rm eV$    & $-4.75~ \rm eV$  \\    
$e_{14}$
  & $0.035 ~\rm  C/m^2  $~\tablenotemark[5]     
& $0.068 ~\rm  C/m^2$~\tablenotemark[5]\\
  $\epsilon_R$         & $12.61$                               
& $12.61$~\tablenotemark[1]             \\

$C_{xxxx}$    & $10.22 \times 10^{11}~ \rm dyne/cm^2$   
&     $12.17 \times 10^{11}~ \rm dyne/cm^2$   \\
$C_{xxyy}$     &$5.76 \times 10^{11} ~\rm dyne/cm^2$      
&       $6.01 \times 10^{11} ~\rm dyne/cm^2$   \\
$C_{xyxy}$     & $4.6 \times 10^{11} ~\rm dyne/cm^2$       
&    $5.82 \times 10^{11} ~\rm dyne/cm^2$   \\
$a$                   & $0.58687 ~\rm nm$                              
 &     $0.56532 ~\rm nm$                                \\

\end{tabular}
\tablenotetext[1]{ Value for InP used.}
\tablenotetext[2]{ Reference\ \cite{dp2}}
\tablenotetext[3]{ Reference\ \cite{DP}.}
\tablenotetext[4]{see text.}
\tablenotetext[5]{ Reference\ \cite{e14}.}

\end{table}

\begin{figure}
\caption{Island geometry.  The distances to the $\{011\}$
 and $\{\bar 111\}$ planes
measured from the center of the base of 
the island  are in the ratio  $d_{011}/d_{\bar 111} \approx 1$.
}
\label{fig1}
\end{figure}

\begin{figure}
\caption{Piezoelectric effect for an island with $h=15 ~ \rm nm$.
 (a) Piezoelectric charge density.
The red surface is the contour of $\rho = +0.001 e/\rm nm^3$, blue 
surface indicates
 $\rho = - 0.001 e/\rm nm^3$.
 (b) Electrostatic potential due to the piezoelectric polarization, $V_p$. 
The red surface is the contour of $V_p = +20 mV$, blue
surface  indicates
 $V_p = -20 mV$.
}
\label{fig2}
\end{figure}

\begin{figure}
\caption{
Band diagrams with inclusion of strain. Scale shown is for $h=15 ~\rm nm$, but
other island sizes are obtained by simply rescaling.
(a) Along $\hat z$ direction, through the center of the island. (b)  Along
the line $x=y$ passing through the base of the island.
}
\label{fig3}
\end{figure}

\begin{figure}
\caption{
Conduction-band wave functions and corresponding energies for $h=15 ~\rm nm$.
 red/blue indicates
positive/negative $\psi$.
}
\label{fig4}
\end{figure}

\begin{figure}
\caption{
Valence-band wave functions for $h=15 ~\rm nm$.
 The contours are of  $\sum_{i}{| \psi_i|^2}$ equal to
$0.01$ of the peak value.  (a) Blue surface
 is the ground state, $A_0$,  yellow surface is the 
highest   state located above the island, $B_0$. (b) blue 
surface is the
excited state $A_n$ immediately above $B_0$ in energy.    
}
\label{fig5}
\end{figure}

\begin{figure}
\caption{
Energies as a function of island size. (a) Electron states.
 (b) Valence band.  Only the first few $A$ states are shown, so
there are missing states between the lowest $A$ state and the
highest $B$ state ($B_0$).}
\label{fig6}
\end{figure}

\begin{figure}
\caption{
Magnitude and polarization dependence of band-to-band
optical matrix elements for an island with $h=15 ~\rm nm$.
$I_x + I_y$ is in equal arbitrary units in all three figures.
(a) Transitions from first six conduction-band states $C_n$ 
to valence-band state $A_0$,
(b) Transitions from first six conduction-band states $C_n$ 
to valence-band state $B_0$,
(c)  Transitions from conduction-band  ground state $C_0$ 
to valence-band states $A_0$...$A_{14}$,$B_0$. Transitions 
spaced less than $0.1 \rm meV$ 
apart have been combined.
}
\label{fig7}
\end{figure}

\end{document}